\documentstyle[aps,preprint]{revtex}
\begin{document}
\draft
\title{Dynamics of a Driven Single Flux Line in Superconductors}
\author{Colin Denniston$^{(1)}$ and Chao Tang$^{(2)}$}
\address{$^{(1)}$Department of Physics, Princeton University, 
Princeton, NJ 08544}
\address{$^{(2)}$NEC Research Institute, 4 Independence Way, Princeton,
NJ 08540}

\date{July 27, 1994}
\maketitle

\begin{abstract}
We study the low temperature dynamics of a single flux line in a bulk type-II
superconductor, driven by a surface current, both near and above the onset
of an instability which sets in at a critical driving.  We found that
above the critical driving, the velocity profile of the flux line
develops a discontinuity.
\end{abstract}

\newpage
\narrowtext

Dynamics of a driven elastic string have attracted much recent attention 
\cite{string,Tang94}.  While most of the work has been focused on the 
interesting physics of pinning-depinning transitions in the case of bulk
driving, the paper by Tang, Feng, and Golubovic \cite{Tang94} studied the
case of a surface-current-driven flux line in a bulk type-II superconductor. 
They found a novel instability of the flux line motion at large driving 
currents.  The instability sets in at a critical driving, where the line 
loses its steady state motion and (presumably) will be stretched longer 
and longer.  Their finding depends crucially on the boundary condition
they use.  Physically, the surface driving current is within a boundary
layer of thickness $\lambda$, where
$\lambda$ is the penetration depth.  The boundary condition used in Ref.
\cite{Tang94} is somewhat equivalent to taking the limit $\lambda
\rightarrow 0$ in a plausible but uncontrolled way.  Since the
instability sets in at or near the boundaries, it is necessary to
examine the situation carefully using a more physical boundary layer.
Also, it is important to see what happens when the driving current is
larger than the critical driving -- a question which can not be
addressed by using the boundary condition in Ref. \cite{Tang94}.

In this paper we analyze the flux motion with the more physical
boundary layer Lorentz driving force.  We first use the method of matching 
asymptotic expansions to study the steady state solutions.  The lowest order 
matching condition justifies the form of the boundary conditions
used in \cite{Tang94} and gives the relation of the driving force to the
current.  We then study, both numerically and analytically, the complete 
equation below and {\it above} the onset of instability.

Let us first derive the equation for the flux line motion which involves
the Lorentz force as a term in the equation, as opposed to just a boundary
condition.  As we will be mostly interested in fairly large driving forces,
we neglect pinning effects. The Lorentz force on a flux line is just 
${\bf F}={1\over c}\int{\bf j} \times {\bf h} ds dA$ where $s$ is the
arclength along the flux line and $dA$ a section of infinitesimal area
transverse to the flux line.  If the applied current, ${\bf j}$, is slowly
varying in the direction {\it transverse} to the line, then the integration
in these coordinates may be carried out to give
\begin{equation}
{\bf F}= {\phi_{0} \over c}\int {\bf j} \times {\bf t}ds,
\label{intLor}
\end{equation}
where $\phi_{0}$ is the flux quantum, and ${\bf t}$ is the unit tangent
vector in the direction of the local magnetic field (arclength is taken to be
increasing in the direction of the magnetic field).  The exact form of the
current depends on the geometry of the sample, however it is known that
the magnitude of the applied current drops exponentially with distance
from the boundary of the sample.

For simplicity we model the dynamics of a single flux line as a two
dimensional problem, defined by its shape function ${\bf r}(s,t)$, or
where the parameterization is well defined, $y(x,t)$ (see Fig.\
\ref{fig1}).  The applied field 
is in the negative x-direction, and the applied current in the negative 
z-direction, thus giving the driving Lorentz force predominantly in the 
y-direction.  The Lorentz force {\it per unit length} is then
\begin{equation}
{d{\bf F} \over ds}= {\phi_{0} \over c} j_{0} \left\{ \exp\left({x-L/2 \over 
\lambda}\right)+\exp\left({-x-L/2 \over \lambda}\right) \right\} {\bf n},
\label{LorFor}
\end{equation}
where $j_{0}$ is the current density at the surface, the sample boundary
is at $x=\pm L/2$, and ${\bf n}={\bf \hat z}\times {\bf t}$ is the local 
unit normal vector of the flux line.

The equation of motion for the flux line becomes
\begin{equation}
\gamma \dot{\bf r} = \left[ \sigma K+{\phi_{0} \over c} j_{0} 
\left\{ \exp\left({x-L/2 \over \lambda}\right)+\exp\left({-x-L/2 \over 
\lambda}\right) \right\} \right] {\bf n}.
\label{reqn}
\end{equation}
The left-hand side, the viscous damping, and the first term on the right,
the normal force due to line tension, are the same as in \cite{Tang94},
and the last term is the Lorentz force, from Eq.~(\ref{LorFor}).  $\gamma$
is the damping coefficient ($\gamma\approx{\phi_{0}^2 \over 2 \pi\xi^2 
c^2\rho_{n}}$ from the Bardeen-Stephen model \cite{Tinkham75}, with $\xi$
the coherence length, and $\rho_{n}$ the normal state resistivity), and
$\sigma$ is the line tension, given approximately by ${H_{c}^2\over 8\pi}
 4 \pi \xi^2 \ln(\kappa)$, with $H_{c}$ the critical field and 
$\kappa\equiv\lambda/\xi$ the Ginzburg-Landau parameter.  $K$ is the
curvature, and we have the relations ${\bf t}=\partial_{s}{\bf r}$ and
$\partial_{s}{\bf t}=K{\bf n}$. 

In cases where the tangent vector never becomes vertical (ie. $\partial 
y/\partial x$ remains finite), Eq.~(\ref{reqn}) can be rewritten 
in terms of the $x$ and $y$ coordinates of ${\bf r}$, now 
reparameterized by $x$.  Note that a displacement of $\Delta n$ in the 
direction of the normal is related to a displacement $\Delta y= \Delta n
 \sqrt{1+(\partial y/ \partial x)^2}$ and the curvature in terms of $x$ 
and $y$ is given by $K=[1+(\partial y/\partial x)^2]^{-3/2} (\partial^2 
y/\partial x^2)$.  Thus we get the equation
\begin{equation}
{\gamma{\partial y \over \partial t}\over\sqrt{1+({\partial y\over
\partial x})^2}}={\sigma{\partial^2 y \over \partial x^2} \over 
\left[ 1+({\partial y\over\partial x})^2 \right]^{{3\over 2}}} + 
{\phi_{0} j_{0} \over c}\left[ \exp\left({x-L/2 \over \lambda}
\right)+\exp\left({-x-L/2 \over \lambda}\right)\right].
\label{xyeqn}
\end{equation}

We now examine the {\it steady state} solutions of Eq. (\ref{xyeqn}).  
Steady state implies  $v=\partial y/\partial t$ is constant, which allows 
us to rewrite (\ref{xyeqn}) as a first order equation in the sine of 
the tangent angle, $\theta$.  Setting $f={\phi_{0} j_{0} \lambda \over c
 \sigma}$ and $w=\sin \theta=\frac{\partial y}{\partial x}/
\sqrt{1+(\partial y/\partial x)^2}$ we get,
\begin{equation}
\lambda {\partial w \over \partial x} = \left({\gamma \over \sigma}v
\right)\lambda \sqrt{1-w^2}-f\left[\exp\left({x-L/2 \over \lambda}
\right)+\exp\left({-x-L/2 \over \lambda}\right)\right].
\label{weqn}
\end{equation}
For $x$ far from the boundaries (i.e. $|x\pm L/2|\gg\lambda$) the driving
term is negligible and the equation becomes
\begin{equation}
{\partial w_{o}\over \partial x}=
\left({\gamma \over \sigma}v\right)\sqrt{1-w_{o}^2},
\end{equation}
which has solution
\begin{equation}
w_{o}=\sin\left({\gamma \over \sigma}v x \right).
\label{woeqn}
\end{equation}
Now we examine the solution near the boundary at $x=L/2$.  The coordinate
appropriate in this region is $\eta=-{x-L/2 \over \lambda}$.  In terms of 
$\eta$, our Eq.~(\ref{weqn}) becomes
\begin{equation}
{\partial w_{i} \over \partial \eta}=-\left( {\gamma\over\sigma}v \right) \lambda\sqrt{1-w_{i}^2}+f e^{-\eta},
\end{equation}
If we expand $w_{i}$ in powers of $\left( {\gamma\over\sigma}v \right)\lambda$
as $w_{i}=w_{i}^{(0)}+\left( {\gamma\over\sigma}v \right)\lambda w_{i}^{(1)}
+...$ we obtain a series of equations for the $w_{i}^{(n)}$.  The first 
two of these equations are
{\samepage
\begin{eqnarray}
{\partial w_{i}^{(0)} \over \partial \eta}& = & f e^{-\eta} \nonumber\\
{\partial w_{i}^{(1)} \over \partial \eta}& = & -\sqrt{1-(w_{i}^{(0)})^2}.
\label{widifeq}
\end{eqnarray}}
Assuming an applied field perpendicular to the boundary, 
these have the solution
{\samepage
\begin{eqnarray}
w_{i}^{(0)}& = &f (1-e^{-\eta}),\nonumber\\
w_{i}^{(1)}& = &\sqrt{1-f^2 (1-e^{-\eta})^2}-1-f \arcsin f(1-e^{-\eta})-\\
           &   &\sqrt{1-f^2} \left\{\log {1-f^2(1-e^{-\eta})+\sqrt{(1-f^2)
(1-f^2(1-e^{-\eta})^2)} \over 1+\sqrt{1-f^2}}+\eta \right\}.
\nonumber
\end{eqnarray}}
If we expand $w_o$ about $x=L/2$ ($\eta=0$) and $w_i$ for large $\eta$ we get
{\samepage
\begin{eqnarray}
w_{o} & \longrightarrow & \sin \left( {\gamma v \lambda L \over 2\sigma} 
\right)-\left( {\gamma v \lambda  \over \sigma} \right) \cos\left( 
{\gamma v \lambda L \over 2\sigma} \right) \eta +...\nonumber\\
w_{i} & \longrightarrow & f+\left( {\gamma v \lambda  \over \sigma} 
\right)\left\{ \sqrt{1-f^2}\left[1-\log{{2(1-f^2) \over 1+\sqrt{1-f^2}}
}\right]-1-f \arcsin f\right\} \label{bd}\\
      &                 & -\sqrt{1-f^2}\quad\eta. \nonumber
\end{eqnarray}}
Matching $w_{o}$ to $w_{i}$ gives, to order ${\gamma v \lambda  \over \sigma}$,
\begin{equation}
v={2 \sigma \over L \gamma} \arcsin\left[f+\left( {\gamma v \lambda  \over
 \sigma} \right)\left\{ \sqrt{1-f^2}\left[1-\log{2(1-f^2) \over 
1+\sqrt{1-f^2}}\right]-1-f 
\arcsin f\right\}\right],\quad (f\le1).
\label{vof}
\end{equation}
Note that the velocity found in \cite{Tang94} is obtained by dropping
the term of order ${\gamma v \lambda  \over \sigma}$ on the right hand
side of (\ref{vof}) and thus is the zeroth order of our asymptotic solution.
This matching procedure is illustrated in Fig.\ \ref{fig2}b which shows 
$w=\sin \theta$ as a function of $x$ for $f=0.9$.  The solid line is a
steady state numerical solution, and the broken and dashed lines show
the inner and outer solutions, respectively.  We see that the outer and
inner solutions agree very well with the numerical result within 
their respective domains of validity.  A composite solution, valid on
the whole domain, can be formed by adding $w_{o}$ and $w_{i}$ and 
subtracting their common part from Eq. (\ref{bd}).  This is 
indistinguishable from the numerical solution in Fig.\ \ref{fig2}b.

The numerical solutions shown in Fig.\ \ref{fig2} were produced from 
solutions of
Eq.~(\ref{reqn}).  This was chosen, rather than Eq.~(\ref{xyeqn})
in x-y coordinates, due to problems arising in the continuity of
$\partial y/\partial x$ and the diverging values of $\partial y/\partial
x$ found at large values of $f$ (see below).  As our equation involves
the position vector, ${\bf r}$, explicitly, we must evolve a set of
vectors $\{ {\bf r}(s) \}$ of positions along the curve (as opposed to,
for instance, following the curvature).  We solve Eq.~(\ref{reqn})
using a finite-difference approach.  The viscous term, $\gamma \dot{\bf
r}$, and the curvature term, $K {\bf n}={\bf r}_{ss}$, can be dealt with
using a Crank-Nicholson type approach for diffusive equations.  This
yields two, x and y, tridiagonal systems linked only at the boundaries.
The Lorentz force in (\ref{reqn}) is then dealt with in an
semi-implicit manner.  The system is remeshed at each time step to
preserve point spacing in regions of high curvature.

For a specific case, we take a sample width, $L$, of $100 \lambda$ and
measure the velocity in the unit of $\sigma/\gamma$.  Fig.\ \ref{fig2}a
shows the
line shapes for $f=0.2$ to $f=1.1$.  We see that the slope remains
fairly small within a penetration depth, $\lambda$ of the boundary,
consistent with the assumptions for Eq.~(\ref{intLor}).  Also, the
analytic solution of Ref. \cite{Tang94} starts to deviate from our
numerical solution near the boundary for large $f$.

Fig.\ \ref{fig3} shows $v$ as a function of $f$.  The crosses are
from steady state numerical solutions, the dashed line is the zero'th
order matching condition from \cite{Tang94}, and the dotted line (for
the region $f \le 1$) is from Eq.~(\ref{vof}).  The zero'th order solution 
suggests that as $f \rightarrow 1$, $v \rightarrow v_{max}=\pi \sigma/\gamma 
L$, implying $\theta \rightarrow \pi/2$; i.e. the flux line ``wets'' the 
boundary.  The more accurate expression, Eq. (\ref{vof}), suggests that 
$v \rightarrow {2 \sigma \over L \gamma} \arcsin (1-({\gamma v \lambda
\over \sigma})(1+\pi/2))< v_{max}$ so the flux line does not become
vertical as $f\rightarrow 1$ (see also the numerical solution in Fig. 
\ref{fig2}).  What then does happen for $f$ greater than one?  As we shall 
see below, the flux line becomes vertical ($\theta \rightarrow \pi/2$) at an
interior point, but not until $f=f^{*}=1.07623$ for our sample case where 
$\lambda/L=0.01$.  Above $f^{*}$ the speed of the flux line develops a 
discontinuity, becoming piecewise constant.

Note that in the above analysis $w_{i}^{(0)}>1$ for $f>1$, so it
can not be extended to the region where $f>1$.  This problem can be
remedied by adjusting the arbitrary constant in $w_{i}^{(0)}$ so
that $w_{i}^{(0)}$ does not exceed one.  This means that $w_{i}^{(0)}$
will no longer satisfy the boundary condition $w_{i}^{(0)}|_{\eta=0}=0$.
We can, however, adjust the constant in $w_{i}^{(1)}$ to compensate for
this discrepancy so that  $w_{i}^{(0)}+({\gamma v \lambda \over
\sigma})w_{i}^{(1)}= 0$ at $\eta=0$.  This results in a solution to
Eq.~(\ref{widifeq}), for $f>1$, of
{\samepage
\begin{eqnarray}
w_{i}^{(0)}& = &1-f e^{-\eta},\nonumber\\
w_{i}^{(1)}& = &({\sigma \over \gamma v \lambda})(f-1)+\sqrt{{2 \over
f}e^{-\eta}-e^{-2\eta}}-\sqrt{{2 \over f}-1}+\\
           &   &2 \arcsin \sqrt{{e^{-\eta} \over 2 f}}-2 \arcsin \sqrt{{1
\over 2 f}}.\nonumber
\end{eqnarray}}
Expanding this $w_{i}$ for large $\eta$ gives
\begin{equation}
\label{wias}
w_{i}\longrightarrow f-\left({\gamma v \lambda \over \sigma}\right)\left[
\sqrt{{2 \over f}-1}+2 \arcsin \sqrt{{1 \over 2 f}}\right].
\end{equation}
Matching this to $w_{o}$ at $x=L/2$ gives, to order ${\gamma v \lambda
\over \sigma}$,
\begin{equation}
v={2 \sigma \over L \gamma} \arcsin\left[f-\left({\gamma v \lambda \over
\sigma}\right)\left\{\sqrt{{2 \over f}-1}+2 \arcsin \sqrt{{1 \over 2 f}}
\right\}\right].
\label{vof2}
\end{equation}
Note that for $f=1$, $v={2 \sigma \over L \gamma} \arcsin (f-({\gamma v
\lambda \over \sigma})(1+\pi/2))$, the same result as taking
$f\rightarrow 1$ in Eq.~(\ref{vof}).  Eq.~(\ref{vof2}) has
only real solutions for $f \le f^{*}=1.07623$ (for $L=100 \lambda$).  It 
suggests that the instability should occur at $f=f^{*}$ where
$v=v_{max}=\pi\sigma/\gamma L$.  Eq.~(\ref{vof2}) is shown
as the continuation of the dotted line for $1<f<f^{*}$ in Fig.\
\ref{fig3}.  For general $\lambda/L$, $f^{*}$ is found as the root
of Eq.~(\ref{vof2}) for $v=v_{max}=\pi\sigma/\gamma L$.  For small 
$\lambda/L$, the case we are interested in, this root is 
\begin{equation}
f^{*} \approx 1+{\lambda \over L}(\pi+\pi^{2}/4)
\label{fstar}
\end{equation}
We see that as $\lambda/L \rightarrow 0$, $f^{*} \rightarrow 1$.

The question now arises as to what happens above $f^{*}$.  Fig.\
\ref{fig4}a
shows the numerical evolution of the flux line shape for $f=1.1$, just
above the transition, and $f=1.5$.  There are two important things to
note in this figure.  First, the flux line is approaching a vertical
asymptote at about $x=40.5$ in what seems to be an asymptotic manner
(i.e. the flux line does not become vertical in a finite amount of time).
Secondly, the portion of the flux line to the boundary side of this
vertical asymptote has a constant shape, implying {\it it is moving with
a constant speed in the y-direction}.  This last observation can be
verified by applying a finite difference approximation to Eq.~(\ref{xyeqn})
to compute $\partial y/\partial t$ for the points on the
flux lines of Fig.\ \ref{fig4}a.  The result of this computation is shown in 
Fig.\ \ref{fig4}b.  We see from this velocity profile that, indeed the speed is
constant in the boundary layer, but that a {\it discontinuity} has developed
in the velocity profile!  The constant speed of the boundary layer can
be deduced as follows.

Eq.~(\ref{wias}) gives the {\it constant} asymptotic value of $w_i$
for large $\eta$.  Above $f^*$ this asymptotic value can only be one
($w_i=\sin \theta\leq 1$), as the flux line becomes vertical.  So
setting (\ref{wias}) to one gives the speed of the inner solution,
$v_i$, for $f>f^*$ as
\begin{equation}
v_i={{2 \sigma \over \gamma L}(f-1) \over \sqrt{{2 \over f}-1}+2
\arcsin\sqrt{{1 \over 2 f}}}.
\end{equation}
This is shown as the dotted line in Fig.\ \ref{fig3} for $f>f^*$.  
Comparison of this speed to the speed obtained 
in the numerical simulations shows excellent agreement.

Now, what about the outer solution, $w_{o}$?  We see from Fig.\
\ref{fig4} that as
time progresses the speed of the outer solution approaches a constant
value and that the position of the discontinuity in the velocity profile
(or the vertical asymptote in Fig.\ \ref{fig4}a) seems to approach a 
fixed value.
The location of the vertical asymptote and the asymptotic (large time)
speed of the inner solution are quite related.  Requiring that
$w_o\rightarrow1$ at the vertical asymptote gives the speed of the outer
solution.

In conclusion, we have studied the flux line motion, in particular the
dynamical instability found in Ref. \cite{Tang94}, using a more physical
boundary layer driving.  The boundary condition used in \cite{Tang94} is
consistent with our zeroth order (in $\lambda$) asymptotic matching.
The analytic solution of Ref. \cite{Tang94} is quantitatively valid for
$f \le 0.8$.  For larger $f$, the deviations both in line shape near the
boundary and in the velocity are significant.  We have shown that the
instability occurs at $f=f^*=1+(\lambda/L)(\pi+\pi^{2}/4)$ where the flux 
line starts to loose a steady state motion.  We have observed numerically 
that above this instability the flux line velocity profile develops a
discontinuity.  This instability has a clear mark on the ``$I-V$''
curve, Fig.\ \ref{fig3}, that is a sharp upward turn at $f^*$.  
As pointed out in Ref. \cite{Tang94}, this instability should also 
occur in dense flux line systems.

\newpage

\begin{figure}
\caption{Sketch of the cross section of the sample.  The driving current
is near the sample surfaces.}
\label{fig1}
\end{figure}

\begin{figure}
\caption{(a) Steady state flux line profiles for, from top to bottom, 
$f$=0.2, 0.4, 0.6, 0.8, 1.0, and 1.1.  (b) Matched asymptotic expansions for 
$f=0.9$: numerical solution (solid line), $w_{o}$ (dashed line), 
and $w_{i}$ (dot-dashed line).}
\label{fig2}
\end{figure}

\begin{figure}
\caption{The velocity of the flux line as a function of driving force:
numerical simulations (crosses), Eqs. (12), (15), and (16) (dotted line),
and Eq. (10) of Ref. [1] which is also the zeroth order of Eq. (12) 
(dashed line).}
\label{fig3}
\end{figure}

\begin{figure}
\caption{(a) The shape of the flux line at different times for $f=1.1$ 
(solid lines) and $f=1.5$ (dotted lines).  The solutions which extend
further down the plot are at later times.  (b)  Velocity profiles of 
the flux lines of (a) at the latest times shown.}
\label{fig4}
\end{figure}


\newpage

\begin{references}
\bibitem{string} M. V. Feigel'man, Sov. Phys. JETP {\bf 58}, 1076
(1983); R. Bruinsma and G. Aeppli, Phys. Rev. Lett. {\bf 52},
1547 (1984); J. Koplik and H. Levine, Phys. Rev. B {\bf 32}, 280 (1985);
T. Nattermann, S. Stepanow, L.-H. Tang, and H. Leschhorn, J. Phys. II 
France {\bf 2}, 1483 (1992); G. Parisi, Europhys. Lett. {\bf 17}, 673 
(1992); K. Sneppen, Phys. Rev. Lett. {\bf 69}, 3539 (1992); O. Narayan 
and D. Fisher, Phys. Rev. B {\bf 48}, 7030 (1993); M. Dong, M. C. 
Marchetti, A. A. Middleton, and V. Vinokur, Phys. Rev. Lett. {\bf 70}, 
662 (1993); Z. Csah\'{o}k, K. Honda, E. Somfai, M. Vicsek, and 
T. Vicsek, Physica A {\bf 200}, 136 (1993); H. A. Makse, A.-L. 
Barab\'{a}si, and H. E. Stanley (unpublished).
\bibitem{Tang94} 
C. Tang, S. Feng, and L. Golubovic, Phys. Rev. Lett. {\bf 72}, 1264
(1994).
\bibitem{Tinkham75} 
M. Tinkham, {\it Introduction to Superconductivity} (McGraw-Hill, New
York,1975), p162.
\end{references}
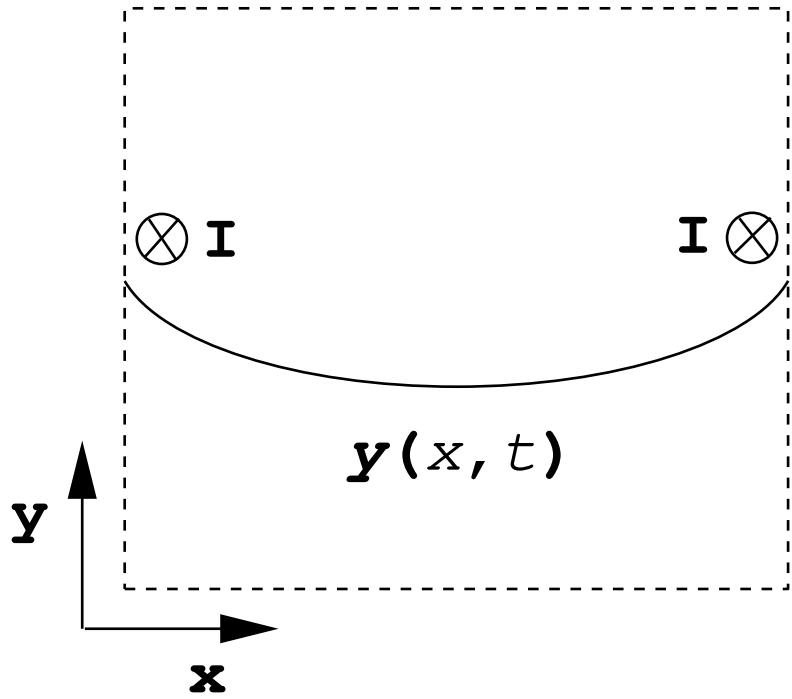
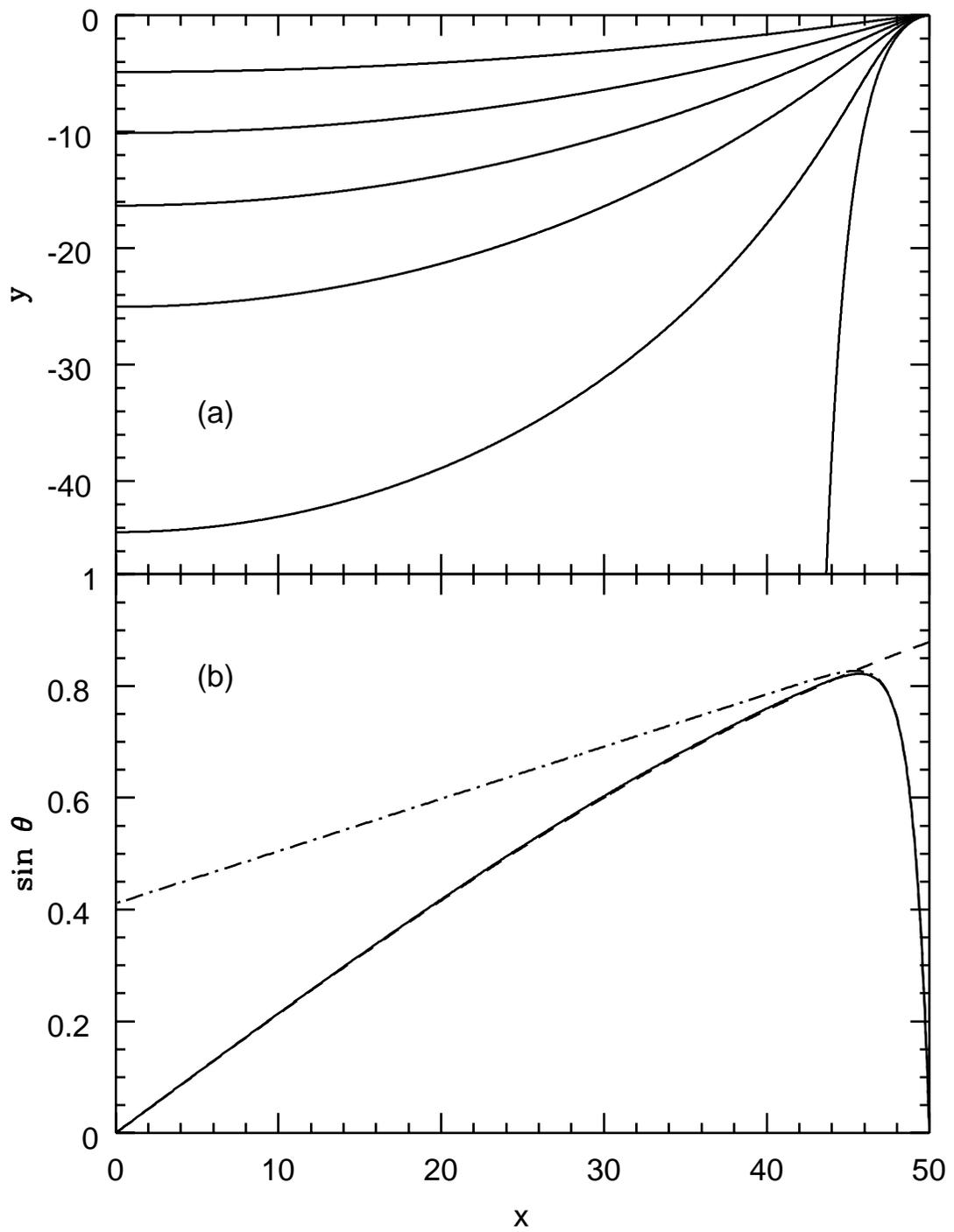
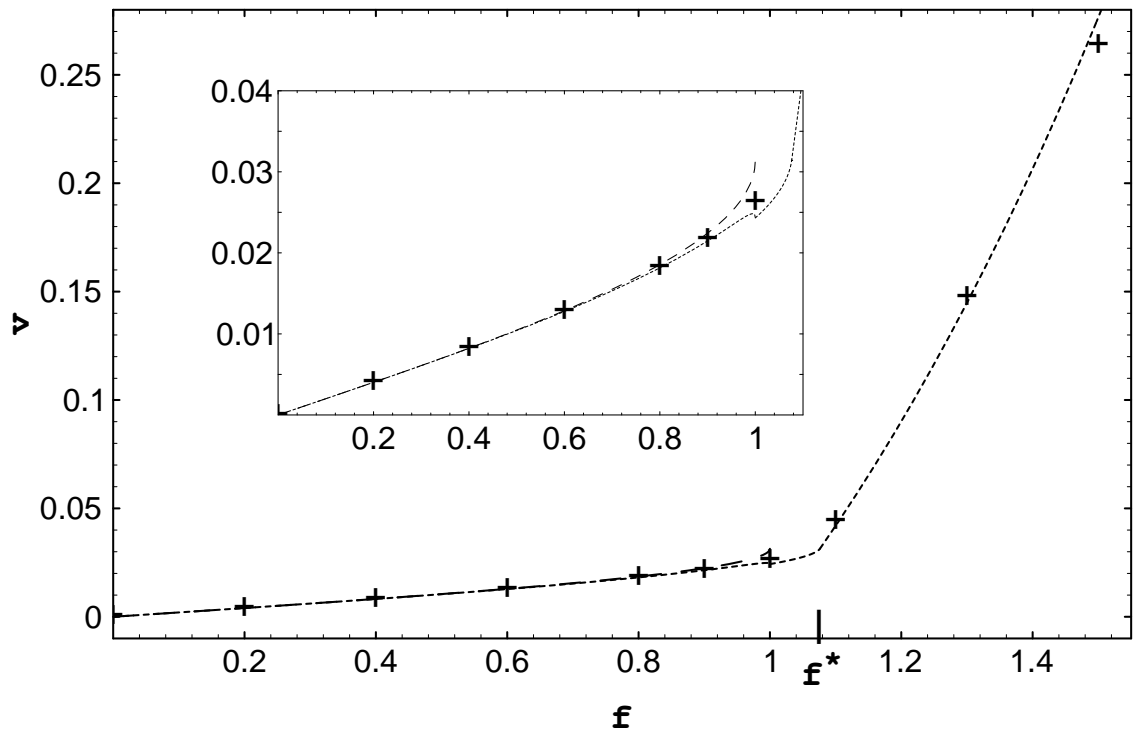
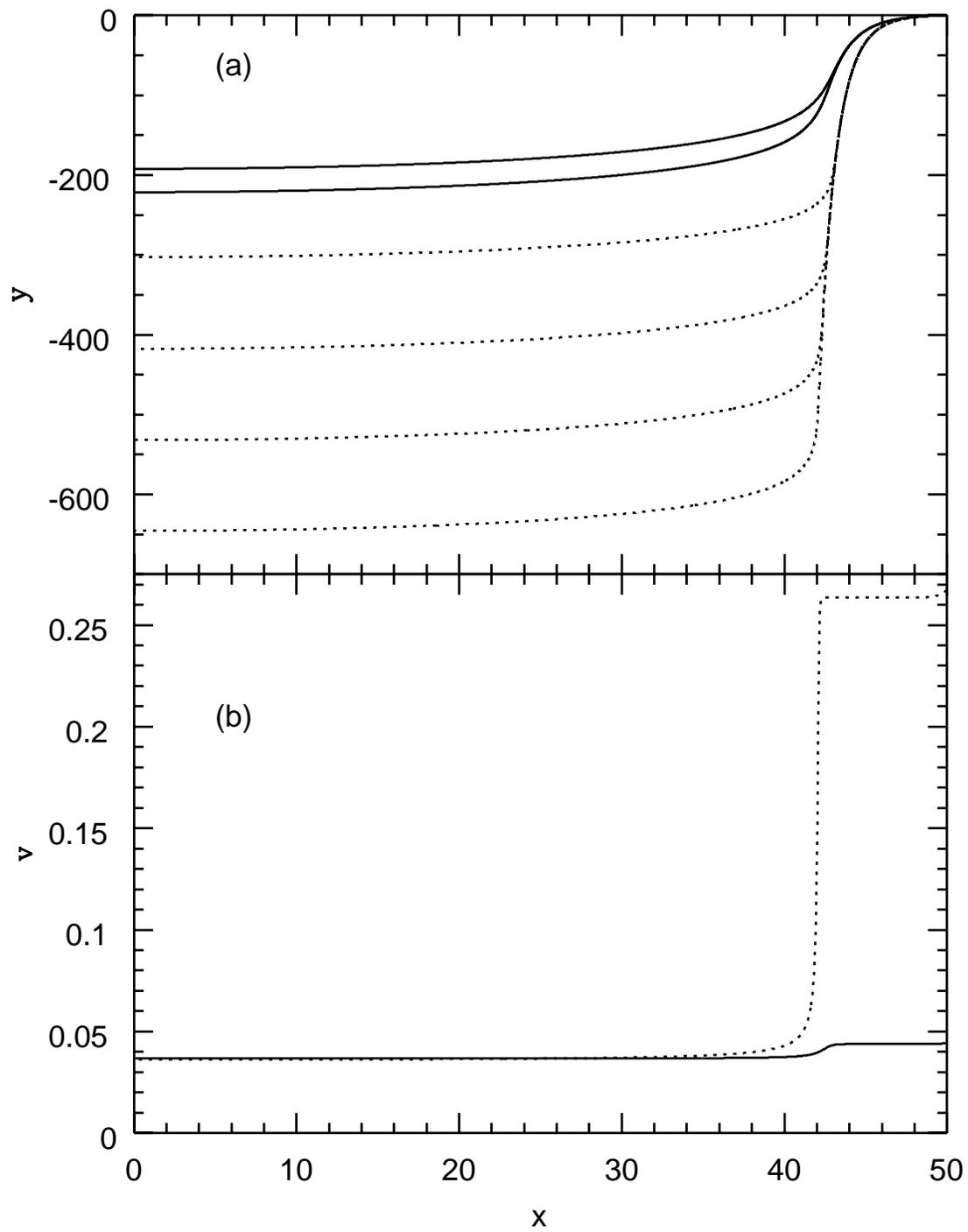
\end{document}